\documentclass[conference]{IEEEtran}
\IEEEoverridecommandlockouts
\usepackage{cite}
\usepackage{amsmath,amssymb,amsfonts}
\usepackage{algorithmic}
\usepackage{graphicx}
\usepackage{textcomp}
\usepackage{xcolor}
\usepackage{here}
\usepackage {booktabs}
\usepackage{subcaption}
\usepackage{url} 

\def\BibTeX{{\rm B\kern-.05em{\sc i\kern-.025em b}\kern-.08em
    T\kern-.1667em\lower.7ex\hbox{E}\kern-.125emX}}
\begin{document}

\title{Investigating Quantitative-Qualitative Topical Preference: A Comparative Study of Early and Late Engagers in Japanese ChatGPT Conversations}

\author{\IEEEauthorblockN{Tomoki Fukuma}
\IEEEauthorblockA{\textit{TDAI Lab Co., Ltd., Japan} \\
tomoki.fukuma@tdailab.com}
\and
\IEEEauthorblockN{Koki Noda}
\IEEEauthorblockA{\textit{TDAI Lab Co., Ltd., Japan} \\
koki.noda@tdailab.com}
\and
\IEEEauthorblockN{Yuta Yamamoto}
\IEEEauthorblockA{\textit{TDAI Lab Co., Ltd., Japan} \\
yuta.yamamoto@tdailab.com}
\and
\IEEEauthorblockN{Takaya Hoshi}
\IEEEauthorblockA{\textit{TDAI Lab Co., Ltd., Japan} \\
takaya.hoshi@tdailab.com}
\and
\IEEEauthorblockN{Yoshiharu Ichikawa}
\IEEEauthorblockA{\textit{Japan Broadcasting Corporation} \\
ichikawa.y-gq@nhk.or.jp}
\and
\IEEEauthorblockN{Kyosuke Kambe}
\IEEEauthorblockA{\textit{Japan Broadcasting Corporation} \\
kambe.k-je@nhk.or.jp}
\and
\IEEEauthorblockN{Yu Masubuchi}
\IEEEauthorblockA{\textit{Japan Broadcasting Corporation} \\
masubuchi.y-lq@nhk.or.jp}
\and
\IEEEauthorblockN{Fujio Toriumi}
\IEEEauthorblockA{\textit{The University of Tokyo} \\
tori@sys.t.u-tokyo.ac.jp}


}

\maketitle

\begin{abstract}
This study investigates engagement patterns related to OpenAI's ChatGPT on Japanese Twitter, focusing on two distinct user groups - early and late engagers, inspired by the Innovation Theory. Early engagers are defined as individuals who initiated conversations about ChatGPT during its early stages, whereas late engagers are those who began participating at a later date. To examine the nature of the conversations, we employ a dual methodology, encompassing both quantitative and qualitative analyses. The quantitative analysis reveals that early engagers often engage with more forward-looking and speculative topics, emphasizing the technological advancements and potential transformative impact of ChatGPT. Conversely, the late engagers intereact more with contemporary topics, focusing on the optimization of existing AI capabilities and considering their inherent limitations. Through our qualitative analysis, we propose a method to measure the proportion of shared or unique viewpoints within topics across both groups. We found that early engagers generally concentrate on a more limited range of perspectives, whereas late engagers exhibit a wider range of viewpoints. Interestingly, a weak correlation was found between the volume of tweets and the diversity of discussed topics in both groups. These findings underscore the importance of identifying semantic bias, rather than relying solely on the volume of tweets, for understanding differences in communication styles between groups within a given topic. Moreover, our versatile dual methodology holds potential for broader applications, such as studying engagement patterns within different user groups, or in contexts beyond ChatGPT.
\end{abstract}


\section{Introduction}

ChatGPT\footnote{\url{https://chat.openai.com/}}, which is a chatbot released by OpenAI in November 2022 has significantly changed the digital landscape and gained notable attention, eliciting diverse responses from users across different communities~\cite{Leiter2023ChatGPTAM}. Understanding those user engagement patterns and interests towards emerging technology is crucial for addressing concerns in the context of filter bubbles which refers to the tendency of individuals to be exposed to information and perspectives that align with their existing beliefs and preferences while being isolated from diverse viewpoints.  

In this study, we aim to examine the comparative engagement patterns with Japanese ChatGPT on Twitter from a bird’s eye view, focusing on two distinct groups, namely \textit{early engagers} and \textit{late engagers}. Early engagers are defined as users who began interacting with ChatGPT shortly after its launch, while late engagers are users who initiated their interactions in a later phase. The categorization into these groups is inspired  by the Innovation Theory~\cite{UBHD2028615}, which posits that early adopters often constitute a risk-taking, socially influential group, whereas the later majority are typically more cautious and less socially active, often avoiding risk.

Existing research on social interactions on Twitter mainly focused on quantitative analysis, focusing on  proportions derived from topic modeling, and evaluating sentiments and stances among specific groups. However, these studies do not take into account the semantic structure within the topics discussed by different groups. In other words, even if two groups share the same volume of tweets, the manner information spreads within these groups may vary. For instance, while one group may predominantly retweet the same tweets, the other group may create and share more original content, resulting in a situation where later groups disseminate more diverse information. To address this gap, our paper conducts qualitative analysis that extract overlapping and unique content parts of what each group was talking about, aiming to gain a comprehensive understanding of their perspectives.  

To investigate these engagement patterns, we have developed a novel methodology that enables both quantitative and qualitative comparative analyses of the discussion topics. The qualitative examination seeks to identify dominant topics for each group, discerning intriguing thematic differences. For instance, our analysis revealed that early engagers tend to focus on technology-centric subjects, while late engagers lean towards socio-cultural dialogues. Moreover, our in-depth analysis of the qualitative differences between early and late engagers revealed that early engagers tend to focus on a more limited perspectives, whereas late engagers exhibit a broader range of viewpoints within the same topic. Furthermore our analysis identified that a high volume of discussion does not necessarily result in semantic diversity within a topic. This underscores the importance of diving beneath the surface level of engagement numbers to understand the actual content and nuances of the discussions happening among users.

In conclusion, our research offers valuable insights into the public discourse surrounding ChatGPT, shedding light on the unique conversational inclinations of early and late engagers. Additionally, the hybrid appriach of quantitative and qualitative methods is versatile enough for broader applications in studying engagement patterns within other user groups and contexts beyond this specific case. Finally, we make our testing code publicly available on GitHub\footnote{\url{https://github.com/TDAILab/semantic_polarization}}.

\section{Related Work}
The advent of ChatGPT has led to an increase in research aimed at exploring public perceptions towards it, with Twitter data as the primary source of information. Haque et al.~\cite{Haque} employed Latent Dirichlet Allocation (LDA) for topic modeling to identify popular topics in ChatGPT-related tweets. Subsequently, they performed sentiment analysis based on these topics. Taecharungroj~\cite{Taecharungroj2023WhatCC} also made use of LDA topic modeling but their dataset was larger, comprising over 200,000 tweets. Their focus was more on identifying the strength and limitations of ChatGPT rather than examining public opinions. In a more encompassing study, Leiter et al.~\cite{leiter} carried out sentiment analysis and classified tweets into 19 pre-defined topics. This was facilitated by a model built upon roBERTa that was fine-tuned explicitly for tweet topic classification\cite{antypas-etal-2022-twitter}. They expanded their analysis to incorporate differences across various languages and over time.

Unlike the aforementioned studies, our research is geared towards a comparative study between groups. We also offer a distinctive approach by calculating the proportion of shared or unique content between groups. 
This innovative method provides insight into mutual understanding and polarization within groups.

To the best of our knowledge, no studies have examined the combination of traditional topic modeling techniques and bias analysis of semantic vectors like our approach, even for other tweet analysis topics such as COVID-19 and the US election. Studies in the field of fairness in machine learning sometimes investigate the bias of semantic vectors between sensitive attributes such as gender or race. Our methodology draws inspiration from Bolukbasi et al.~\cite{10.5555/3157382.3157584}, who examined biases in word embeddings between genders. They aimed to measure the gender bias of word representations in Word2Vec and GloVe by calculating the projections into principal components of the differences of embeddings of a list of male and female pairs,a concept they termed "gender direction". Our approach is similar in terms of mapping tweets into a dimension that separates early and late engagers and analyze the distributional bias concerning these user groups.

\begin{table}[t]
\caption{Statistical description of Tweet Dataset}
    \label{table1}
    \centering
        \begin{tabular}{cc}
        \hline Attribute & Detail \\
        \hline data range & Nov 30, 2022 to Feb 20, 2023  \\
        \# of users & 255,569 \\
        \# of tweets including RT & 811,050 \\
        \# of tweets excluding RT & 234,148 \\
        language & Japanse \\
        \hline
        \end{tabular}
    
\end{table}

\begin{figure}
    \centering
    \includegraphics[width=\linewidth]{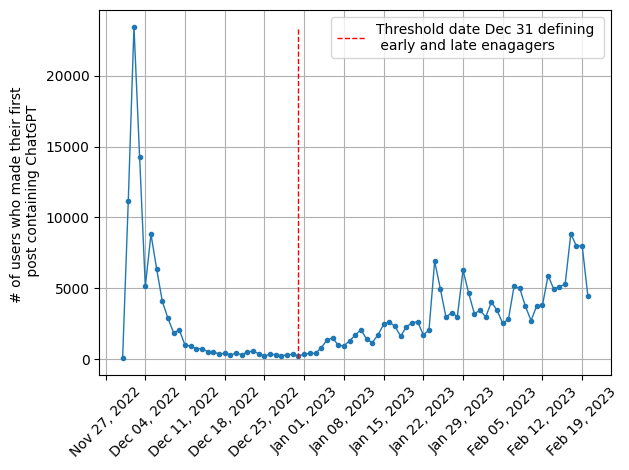}
    \caption{Trends in the daily number of new users who made their first post containing ChatGPT. Early and late engagers are defined based on the threshold set on December 31, corresponding to the minimum value of the seven day average represented by the red dotted line.}
    \label{fig1}
\end{figure}

\begin{table}[tp]
\centering
\caption{Statistical Description of Dataset (Dec 31, 2022 to Feb 20, 2023) Comparing Early and Late Engagers}
\label{tab:engagement_stats}
\begin{tabular}{c c c}
\hline
Metric & Early Engagers & Late Engagers \\
\hline
\# of users & 36,736 & 166,394 \\
\# of tweets including RT & 242,073 & 355,622 \\
\# of tweets excluding RT & 56,568 & 131,170 \\
\hline
\end{tabular}
\end{table}

\section{Data Collection}

\subsection{Source of Data and Collection Technique}
Twitter was selected as the primary data source for this study due to its widespread popularity and utilization as a platform for social interaction.  We originally collected public Japanese tweets that were posted between November 30, 2022, and February 20, 2023, and mentioned the term "ChatGPT". This extraction was facilitated through the Twitter API v2, which is specifically designed for Academic Research. Table \ref{table1} provides a statistical summary of the assembled dataset.

\subsection{Differentiation between Early and Late Engagers}
We aimed to apply the principles of innovation theory in order to distinguish between early and late engagers of ChatGPT.
This theory helps categorize users into several distinct groups based on the timeline of their adoption or involvement with a new product, service, or technology. Specifically, it designates about 2.5\% of the total population as innovators, around 13.5\% as early engagers, 34\% as early majority, another 34\% as late majority, and roughly 16\% as late adopters.

Figure \ref{fig1} provides an illustration of the number of users posting their first tweet about ChatGPT, where the x-axis represents the timeline, and the y-axis indicates the number of users. The ongoing evolution and growth of the ChatGPT topic as depicted in this figure suggest that these precise percentages may not be suitable. In response, we look at the graph of user volumes, noticing an initial spike in user engagement, a subsequent calming period around December, followed by a reacceleration in mid-January when a large number of users started to participate. We interpret these inflection points as "chasms," reflecting the concept from innovation theory, indicating a substantial gap between the innovators or early adopters and the remainder of the users. We thus divided the users into groups at these specific junctures.

To set the threshold for classification, we calculate the average of number of new users over the preceding seven days and selected the lowest count as the threshold. As a result, users who tweeted before December 31, 2022 were classified as early engagers, while those who started tweeting after this date were categorized as the late engagers. Table \ref{tab:engagement_stats} also provides a statistical summary of the assembled dataset on or after December 31, 2022.

\section{Topic Modeling Approach}\label{topicmodeling}
Our study aims to compare engagement patterns between two groups. For this purpose, we firstly adopt a topic modeling to categorize each tweet according to specific themes. Topic modeling can be generally classified into two main types: Bayesian probabilistic topic models (BPTMs), such as Latent Dirichlet Allocation (LDA)\cite{944937}, and clustering-based topic models (CBTMs), including models like BERTopic \cite{grootendorst2022bertopic}. Recent studies indicate that CBTMs outperform BPTMs \cite{zhang-etal-2022-neural}, producing more coherent and diverse topics while requiring fewer computational resources and less time. The subsequent subsections outline the specifics of our implementation.

\subsection{Extracting Semantic Features via Text Embeddings}\label{embedding}
The first step involves extracting semantic features from the tweets using text embeddings. We filter out retweets and preprocess the text by removing mentions and URLs. OpenAI's \textit{text-embedding-ada-002}, is then utilized to transform the cleaned text of each tweet into a 1536-dimensional sentence embedding.

To mitigate the Curse of Dimensionality \cite{Aggarwal01onthe}, which could adversely affect clustering results, we reduce the dimensions of the sentence embeddings. Based on recommendations by \cite{zhang-etal-2022-neural}, we employ Uniform Manifold Approximation and Projection (UMAP) \cite{2018arXivUMAP} to bring down the dimensions from 1536 to 5.

\subsection{Density-Based Clustering and Hyperparameter Tuning}
We use Hierarchical Density-Based Spatial Clustering of Applications with Noise (HDBSCAN) for clustering tweets into topics. HDBSCAN is chosen for its efficacy in grouping similar topics and its ability to discern noise clusters, which contain posts that are not closely related to any particular topic.

For hyperparameter tuning, we use the Density-Based Clustering Validation (DBCV) index \cite{MoulaviJCZS14} to measure clustering quality. The DBCV index evaluates the density connectivity between pairs of data points. We aim to optimize two important parameters of HDBSCAN as implemented by scikit-learn\footnote{\url{https://scikit-learn.org/stable/}}:\textit{min\_cluster\_size} and \textit{min\_samples}. The former specifies the smallest size that HDBSCAN recognizes as a cluster, whereas the latter determines the stringency in defining clusters. We experimented with values [10, 25, 50, 100, 200] for both parameters.

After optimization, the hyperparameters were set to \textit{min\_samples} = 10 and \textit{min\_cluster\_size} = 25, resulting in 390 distinct topics, along with one noise cluster containing tweets not closely associated with any specific topic.

\begin{figure}[tp]
    \centering
    \includegraphics[width=\linewidth]{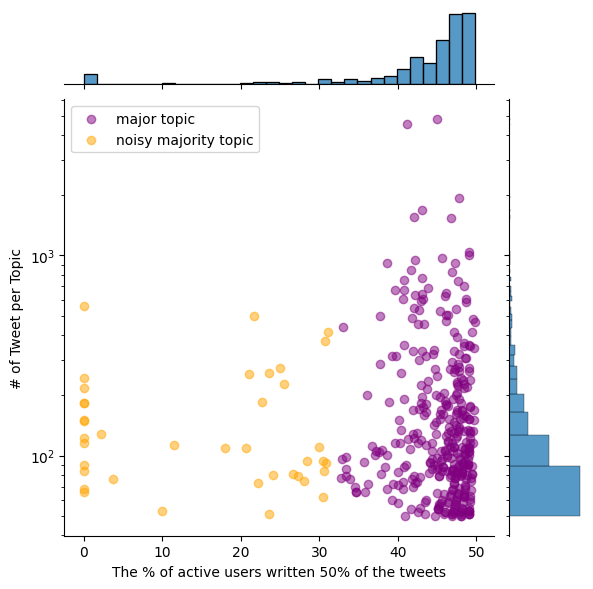}
    \caption{This scatter plot shows topics as individual points, with the y-axis representing the size of the topic and the x-axis indicating the proportion of users responsible for creating half of the content. Topics colored in orange are identified as noisy majority topics using the IQR method and are excluded from further analysis.}
    \label{fig:half}
\end{figure}

\subsection{Topic Filtering}
Once the HDBSCAN algorithm is used to extract topics, we turn our attention to refining the dataset. We observe that some topics are heavily dominated by a small subset of users. We decide to exclude such topics, as they may not be representative topics of the broader community. To identify these topics, we calculate a metric termed "user half line" inspired from~\cite{10.1145/3209581}, which measures the proportion of users contributing to 50\% of the content within a topic. This metric provides a quantitative indication of the concentration of content generation, which may hint at dominant voices or coordinated content dissemination.

Figure \ref{fig:half} visually presents our analysis. The x-axis represents the proportion of users contributing to half of the content, while the y-axis shows the number of tweets in each topic. We use the Interquartile Range (IQR) method with a multiplier of 1.5 to pinpoint outlier topics. Consequently, topics with a user half line of 32.8\% or less are excluded leaving us with a refined dataset containing 351 topics.

\begin{figure}
    \centering
    \includegraphics[width=\linewidth]{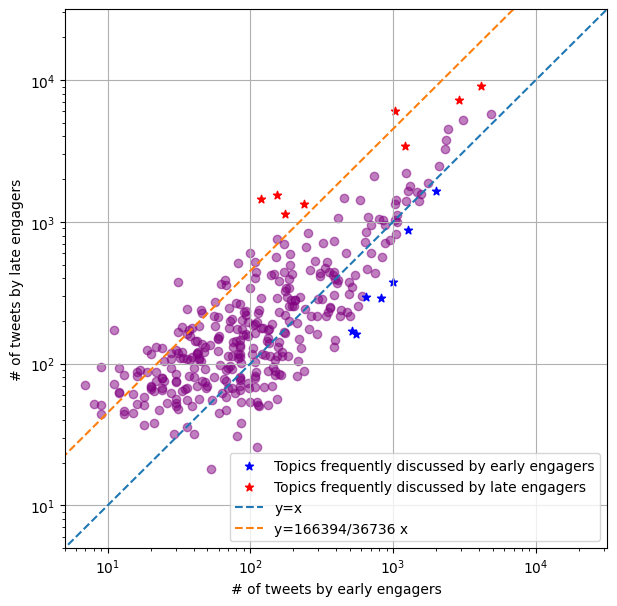}
    \caption{Scatter plot visually represents the tweet activity per person, with each point indicating a specific topic. The x-axis represents the number of tweets per person of early engagers, while the y-axis represents the late engagers. The scatter plot further highlights distinct topics, denoted by red and blue markers, which exhibit notable differences in terms of topic proportions between the two groups.}
    \label{scatter}
\end{figure}

\begin{table*}[]
\centering
\caption{Summaries of topics with particularly different percentages spoken about in Early and Late engagers}
\label{summary}
\begin{tabular}{|l|p{16cm}|}
\hline
\textbf{Dominant} & \textbf{Summary}                                                                                                                                                                                                                                                                              \\ \hline
early             & DeepMind plans to release a chatbot, Sparrow, in 2023, which is speculated to challenge ChatGPT and Google's market dominance with features like misinformation mitigation.                                                                                                                    \\ \hline
early             & A new AI tool, "qqbot", has been integrated into an IDE, providing code conversations and advice on functions and test codes, ushering a new era in software development.                                                                                                                      \\ \hline
early             & Combining the abstract reasoning of ChatGPT and the precise computational knowledge of Wolfram Alpha could create a powerful tool for accurate and wide-ranging question answering.                                                                                                          \\ \hline
early             & ChatGPT’s ability to provide accurate outputs is criticized, and while intended to reduce the knowledge gap, it is recommended to use it with care due to potential output inaccuracies.                                                                                                      \\ \hline
early             & There is surprise and interest surrounding the revelation that ChatGPT can handle tasks previously believed to require logical reasoning.                                                                                                                                                       \\ \hline
early             & The context information is in Japanese and cannot be summarized without understanding the content.                                                                                                                                                                                            \\ \hline
early             & ChatGPT uses a new version of GPT-3 called GPT-3.5 for natural language responses, with anticipation for the expected improved performance of GPT-4 in 2023.                                                                                                                                  \\ \hline
late              & A range of personalized academic writing services are advertised, claiming to assist in various exams and academic writing tasks such as essays and dissertations.                                                                                                                             \\ \hline
late              & AI is used to create an "Automatic Care Report Generator" on Google Sheets, integrating ChatGPT into Google Docs and Sheets for research automation, information gathering, and streamlining administrative tasks.                                                                                \\ \hline
late              & The integration of AI with Notion tool generates infinite articles, providing capabilities like idea generation with ChatGPT, and fact-checking is essential for AI-generated content.                                                                                                        \\ \hline
late              & Various discussions about the capabilities and applications of ChatGPT occur, with differing user experiences and opinions shared online.                                                                                                                                                       \\ \hline
late              & A collection of tweets and headlines discuss a variety of topics, including cryptocurrency news, AI digital tokens, and the potential role of ChatGPT in the cryptocurrency market.                                                                                                             \\ \hline
late              & While ChatGPT struggles with accurate numerical answers, it is useful for those weak in IT and can assist in expressing desires and thoughts, highlighting the evolving role and limitations of AI.                                                                                            \\ \hline
late              & The potential of ChatGPT to replace search engines is discussed, exploring its use in content creation and potential impact on the demand for web articles.                                                                                                                                   \\ \hline
late              & Various uses and limitations of ChatGPT are discussed on Twitter, including language learning, blogging, coding, and information gathering.                                                                                                                                                    \\ \hline
late              & People are exploring ChatGPT's utility for diverse purposes including joke making, prospecting, cold emailing, and more, indicating the AI tool's broad applications and growing popularity.                                                                                                  \\ \hline
\end{tabular}
\end{table*}

\section{Quantitative Analysis in Topic between Early and Late Engagers}

In this section, our objective is to address the research question: \textbf{RQ1. Which topics are prominently discussed by early and late engagers?}

Figure \ref{scatter} presents a scatter plot where each point represents a topic, and its coordinates indicate the number of tweets by early and late engagers. Points located above the line $y=x$ signify topics that late engagers discuss more frequently. Our analysis reveals that 88\% of the topics have a higher number of tweets by early engagers compared to late engagers.

Additionally, we included a line defined by the equation $y = \frac{166349}{36736}x$ in the scatter plot, where the denominator represents the number of late engagers, while the numerator represents the number of early engagers. Points situated below this line indicate that the average number of tweets per person is greater among late engagers than early engagers. Notably, 80\% of the topics fall below this line, suggesting that individuals within the early engagers' group have higher engagement levels per topic.

To identify specific topics that were particularly prominent in each group, we normalize the volume of discussion for each topic based on the number of tweets made by both early and late engagers. We then compute the difference in the proportion of tweets between the two groups using the following formula:

\begin{equation}\label{eq1}
 \frac{1}{2} p_i  \log\left(\frac{p_i}{m_i}\right) + \frac{1}{2} q_i  \log\left(\frac{q_i}{m_i}\right) 
\end{equation}
where $p_i$ is the probability of early engagers discussing topic $i$, $q_i$ is the probability of late engagers discussing the same topic, and $m_i$ is the average topic probability between early and late engagers. This metric is a component of the Jensen-Shannon Divergence~\cite{FugledeTopsoe}.

Due to limited space, we highlight and analyze 16 distinctive topics that are identified as outliers using the IQR method with multiplier of 4. A summary of these topics, generated with a assistance of ChatGPT, is provided in Table~\ref{summary}.

The early engagers appears more invested in the technical potential and future prospects of AI and its development. They are interested in the upcoming versions, possible integrations, and anticipated enhancements in AI. This includes looking at the merging of AI capabilities for improved output, plans of future releases by major tech players, and breakthroughs in integrating AI into different environments, such as IDEs. This perspective indicates an anticipatory stance towards technology, focusing on its development, potential, and its ability to disrupt existing paradigms.

The late engagers, however, seems to focus on the current practical applications, real-world usage, and the observable implications of AI technologies. This includes discussions about their integration into existing platforms for various applications, the pros and cons of these tools in their current state, and their impacts on different fields such as academic writing, content creation, and software development. The "late" group also emphasizes user experiences, pragmatic evaluations, and potential cautionary aspects of AI tools.

Thus, the early engagers embodies a more forward-looking, speculative viewpoint, focusing on technological progression and its potential transformative impact. The late engagers offers a more contemporary and application-focused perspective, centered on maximizing the current capabilities of AI technologies while navigating their limitations. In conclusion, this experimental result resonates well with the innovation theory proposition that early engagers tend to focus on technology-centric subjects, while late engagers lean towards socio-cultural dialogues. The complementarity of these viewpoints is crucial in driving both the evolution of AI technologies and their effective integration into practical applications.

\begin{figure*}
    \centering
    \includegraphics[width=.88\linewidth]{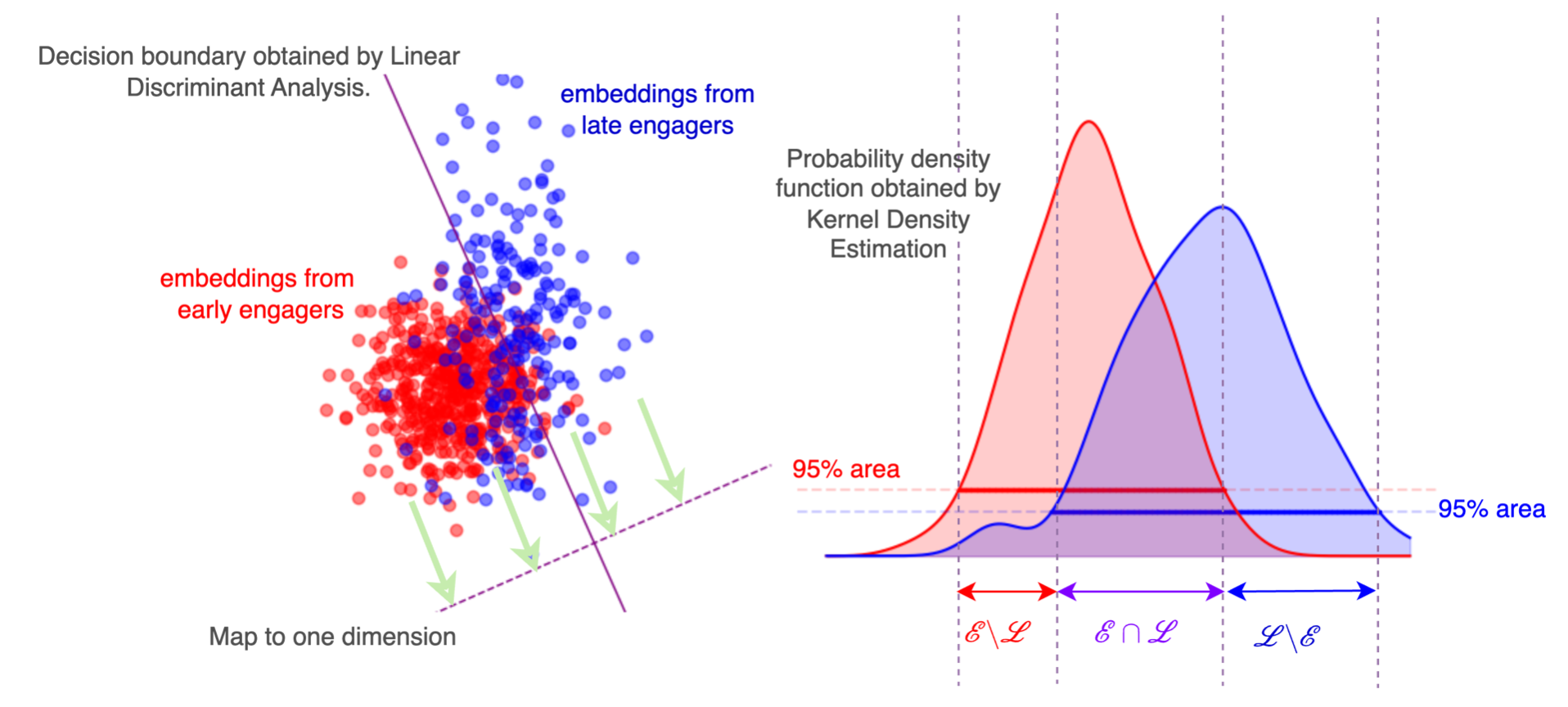}
    \caption{Overview of our proposed method for comparing semantic bias}
    \label{fig:diagram}
\end{figure*}

\begin{figure}
    \centering
    \includegraphics[width=\linewidth]{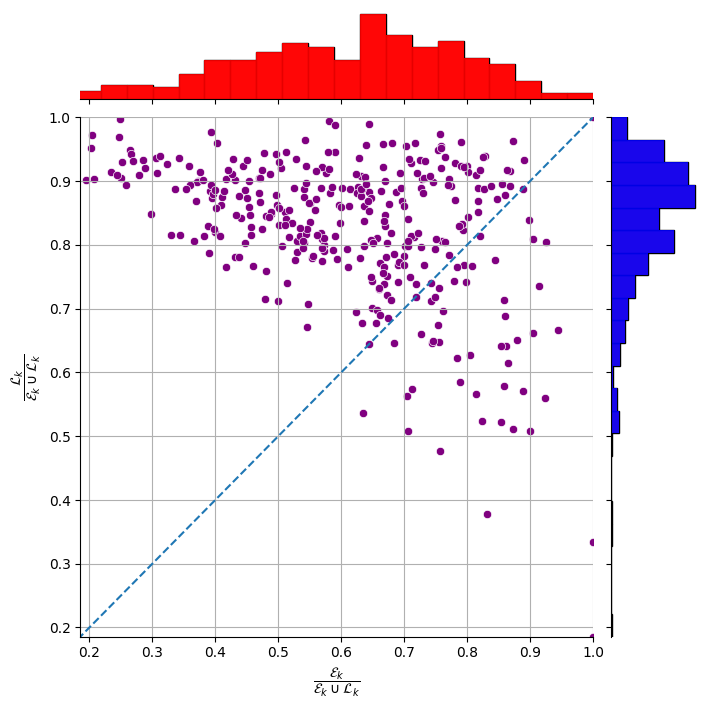}
    \caption{Scatter plot demonstrates a collection of data points, with each point representing a specific topic. The x-axis depicts the magnitude of early engagers, labeled as $\frac{\mathcal{E}_k}{\mathcal{E}_k \cup \mathcal{L}_k }$, while the y-axis portrays the magnitude of late engagers, labeled as $\frac{\mathcal{L}_k}{\mathcal{E}_k \cup \mathcal{L}_k }$.}
    \label{fig:scatter2}
\end{figure}


\begin{figure*}[t]
    \centering
    \begin{subfigure}[b]{0.47\linewidth}
        \includegraphics[width=\linewidth]{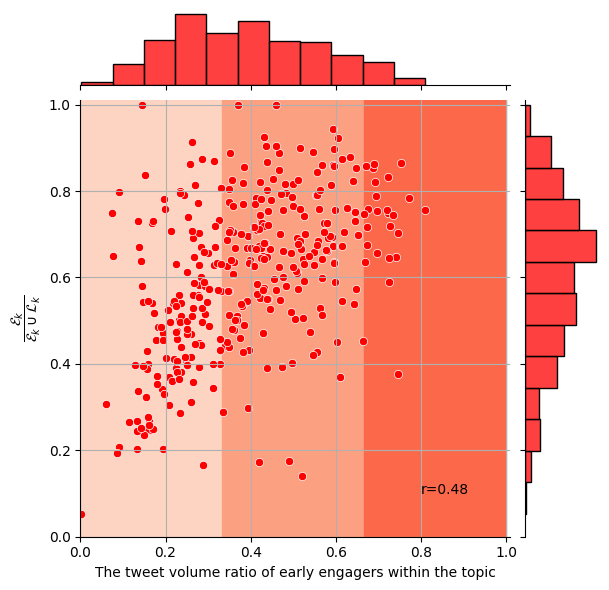}
        \caption{}
        \label{fig:reds}
    \end{subfigure}
    \hfill 
    \begin{subfigure}[b]{0.47\linewidth}
        \includegraphics[width=\linewidth]{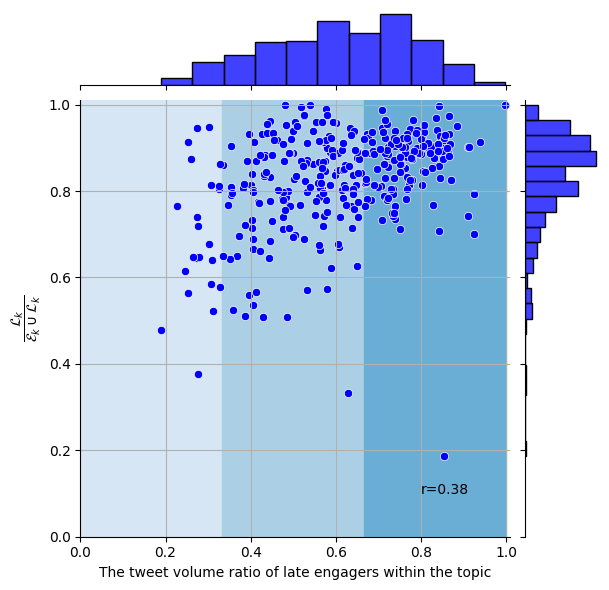}
        \caption{}
        \label{fig:blues}
    \end{subfigure}
    \caption{Scatter plot demonstrates a collection of data points with each point representing a specific topic. The x-axis depicts the ratio of tweet volume of early (a) and late (b) engagers, while the y-axis portrays the $\frac{\mathcal{E}_k}{\mathcal{E}_k \cup \mathcal{L}_k }$ (a) and $\frac{\mathcal{L}_k}{\mathcal{E}_k \cup \mathcal{L}_k }$.}
    \label{fig:redandblue}
\end{figure*}

\begin{figure*}
    \centering
    \begin{subfigure}[b]{0.32\linewidth}
        \includegraphics[width=\linewidth]{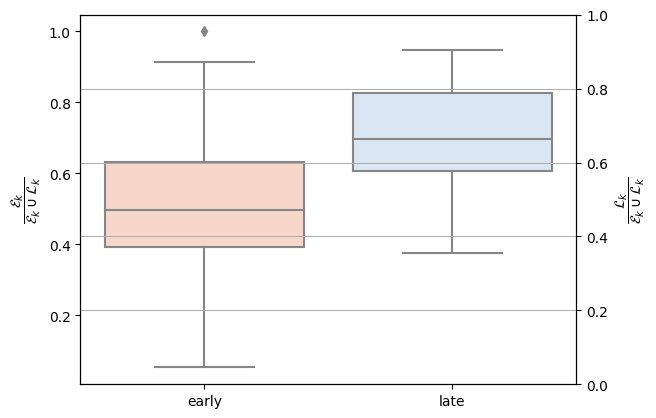}
        \caption{}
        \label{fig:box1}
    \end{subfigure}
    \hfill 
    \begin{subfigure}[b]{0.32\linewidth}
        \includegraphics[width=\linewidth]{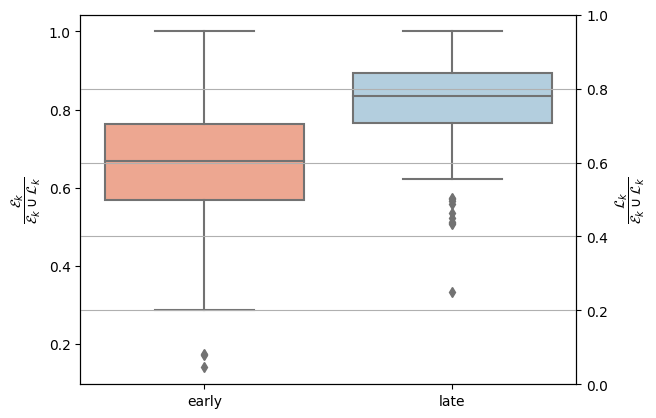}
        \caption{}
        \label{fig:box2}
    \end{subfigure}
    \begin{subfigure}[b]{0.32\linewidth}
        \includegraphics[width=\linewidth]{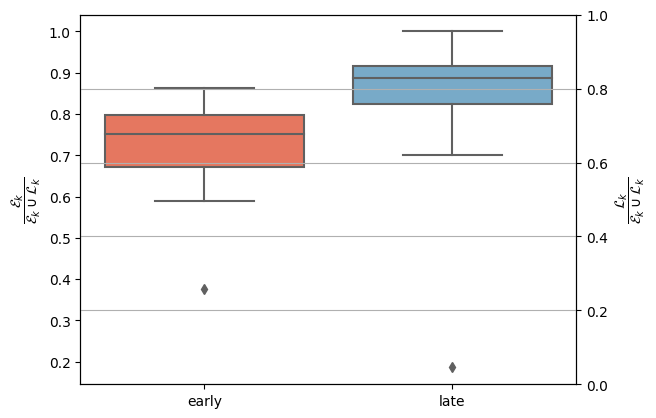}
        \caption{}
        \label{fig:box3}
    \end{subfigure}
    \caption{Box plot showing comparisons between $\frac{\mathcal{E}_k}{\mathcal{E}_k \cup \mathcal{L}_k }$ and $\frac{\mathcal{L}_k}{\mathcal{E}_k \cup \mathcal{L}_k }$ in each group obtained by stratified analysis colored in Figure \ref{fig:redandblue}.}
    \label{fig:box}
\end{figure*}

\section{Quantative Difference within Topics}
In the previous section, we observed that early and late engagers discuss various topics in disparate proportions. This finding prompted an examination of whether these two groups also interact differently within the same topics. 

In the following we aim to answer the following sub-research questions.
\begin{itemize}
    \item \textbf{RQ2-1. Which early or late engagers speaks more broadly on a topic?}
    \item \textbf{RQ2-2. Is there a relationship  between the tweet volume and breadth of topic?}
    \item \textbf{RQ2-3. When the bias in the amount of speech is excluded, which speaks more broadly, the early or late enagers?}
\end{itemize}


\subsection{Semantic Bias Measurement Procedure}

In our study, understanding the differences in the content discussed by two groups is crucial. To measure the semantic polarity between the content discussed by these groups, we have employed a methodology that consists of four steps. This methodology allows us to quantify the extent of divergence in discussions and the areas of overlap or uniqueness. Figure \ref{fig:diagram} provides a graphical representation of the steps involved in this procedure.

\begin{enumerate}
    \item \textbf{Text Embedding:} The first step involves encoding the textual content into a more suitable representation for analysis. By transforming the text into embeddings, as detailed in Section \ref{embedding}, we can leverage the rich information captured in the embeddings for further analysis.
    
    \item \textbf{Linear Discriminant Analysis (LDA):} Once the text is represented as embeddings, we apply Linear Discriminant Analysis (LDA) to map the embeddings of each topic into a single dimension. LDA is particularly advantageous because it seeks to maximize the variance between groups while minimizing the variance within groups. In the context of our study, this effectively translates to identifying perspectives or dimensions where the discussions by the two groups are most divergent. This proceess is similar to the gender direction mapping by previous studies \cite{10.5555/3157382.3157584} measuring embedding bias between men and women, obtained by principal components of differences of embeddings of a list of male and female pairs.

    \item \textbf{Kernel Density Estimation (KDE):} After obtaining the one-dimensional embeddings for each topic through LDA, we employ Kernel Density Estimation (KDE) to estimate the probability density function of each topic. KDE helps in understanding the distribution of discussions for each group across the identified dimension. For this, we set a hyperparameter at 95\% to define the regions for calculating the areas $\mathcal{E}_k$ and $\mathcal{L}_k$, where $\mathcal{E}$ and $\mathcal{L}$ represent the areas of early and late engagers respectively, and $k$ denotes the topic id. The selection method for the bandwidth in KDE is determined through cross-validation \cite{10.1006/jmva.1997.1659}.

    \item \textbf{Ratio of Intersection:} With the probability density functions obtained through KDE, we can now calculate the areas where both groups have discussions ($\mathcal{E}_k\cap \mathcal{L}_k$), areas where discussions are exclusive to early enagagers ($\mathcal{E}_k \backslash \mathcal{L}_k$), and areas where discussions are exclusive to late engagers ($\mathcal{E}_k \backslash \mathcal{L}_k$). This step allows us to quantify the extent of commonality and divergence in the content discussed by the two groups within each topic. 
    
\end{enumerate}

By executing these steps, our methodology provides a systematic and quantitative approach to analyze the differences and similarities in the semantic content of discussions between two groups. This helps in gaining insights into the nature of their communication and identifying the unique and shared aspects of their discussions across different topics.

\subsection{Analyzing Topic Breadth Among Early and Late Engagers}
\label{rq21}
To answer the sub-research question: \textbf{RQ2-1. Do early or late engagers discuss topics more diversely within topics?}, we examined the range of topics brought by each group. 

Figure \ref{fig:scatter2}, the x-axis represents the proportion of areas covered by the early engagers in each topic($\frac{\mathcal{E}_k}{\mathcal{E}_k \cup \mathcal{L}_k }$), while the y-axis shows the same for the late engagers ($\frac{\mathcal{L}_k}{\mathcal{E}_k \cup \mathcal{L}_k }$). This scatter plot allows for a direct comparison of the diversity of topics discussed by both groups.

The mean value for $\frac{\mathcal{E}_k}{\mathcal{E}_k \cup \mathcal{L}_k }$ was found to be 0.6, while the mean value for $\frac{\mathcal{L}_k}{\mathcal{E}_k \cup \mathcal{L}_k }$ was 0.82. A Welch's t-test was conducted on these sets of data, and the results confirmed a significant difference between the two ($p < 0.0001$). This indicates that late engagers generally encompass a more diverse set of subtopics or angles within the primary topics identified in Section \ref{topicmodeling}. Conversely, early engagers seem to be more focused or limited in their discussion within the topics.

\subsection{Comparison between the tweet volume and breadth of topic}

In the previous section, we observed that the late engagers had a more extensive coverage of the topic compared to the early engagers. One of the potential explanations for this could be seen in Figure \ref{fig:half}, which illustrates that late engagers also had a higher volume of tweets per topic than the early engagers. To investigate whether there is a correlation between the breadth of the topics covered and the volume of tweets, a comparative analysis was conducted.

Sub-research question \textbf{RQ2-2 Is there a relationship  between the tweet volume and breadth of topic?}, was explored by plotting the relationship between them in Figure \ref{fig:reds} and \ref{fig:blues}. The x-axis of Figure \ref{fig:reds} shows the ratio of the tweet volume of the early engagers per topic, while the y-axis represents the ratio of covered area by the early engagers ($\frac{\mathcal{E}_k}{\mathcal{E}_k \cup \mathcal{L}_k }$). With a correlation coefficient of 0.48, a relatively weak positive correlation exists between the two factors. Similarly, Figure \ref{fig:blues} represents the late engagers’ data.The correlation coefficient for this set is slightly lower at 0.37, again suggesting a weak correlation.

The identified weak correlations suggest that the broader coverage within topics by late engagers in Section \ref{rq21} cannot be solely attributed to the large number of tweets. The content they share is also a significant factor. These findings highlight the need to consider the semantic nuances in tweet content. Therefore, future research should go beyond tweet volumes and explore the semantic aspects and biases of the content.

\subsection{Stratified Analysis of Topic Breadth by Early and Late Engagers}

In the previous subsection, we discovered that it is essential to minimize the bias caused by differences in tweet volume in order to determine whether early or late engagers participate more in discussions with a broad range of meaning. This subsection answers \textbf{RQ2-3. When the bias in the amount of tweets is excluded, which group speaks more broadly?}

To achieve this, we employ a stratified analysis technique, allowing us to observe the groups within partitions that are more homogenous in terms of the ratio of tweet volume. This approach ensures a more unbiased comparison between the two groups.

Upon manual inspection of Figure \ref{fig:reds} and Figure \ref{fig:blues}, the dataset was divided into three strata based on the percentage of the amount of tweets each group accounted for in a given topic. The strata were formed as follows: up to 1/3, between 1/3 and 2/3, and above 2/3 of the ratio which are colored in Fig\ref{fig:redandblue}.

The results of this stratified analysis are depicted in Figure \ref{fig:box} (a) through (c). In these figures, the x-axis represents the percentage of the amount of tweets that each group accounted for, while the y-axis shows the distribution of $\frac{\mathcal{E}_k}{\mathcal{E}_k \cup \mathcal{L}_k }$ and $\frac{\mathcal{L}_k}{\mathcal{E}_k \cup \mathcal{L}_k }$ for topics assigned to each group. 

A key observation from the results is that, across all strata, the average value of $\frac{\mathcal{L}_k}{\mathcal{E}_k \cup \mathcal{L}_k }$ is consistently higher than the average value of $\frac{\mathcal{E}_k}{\mathcal{E}_k \cup \mathcal{L}_k }$. Furthermore, statistical tests reveal that the differences are significant across all strata ($p < 0.0001$). 

This implies that, even when controlling for tweet volume, late engagers still exhibit a broader range of discussion within the topics. This difference in breadth of topics can be an important consideration in understanding the nature of discussions and the diversity of perspectives presented by different groups.

\section{Conclusion}
This study explored the discourse about ChatGPT within the Japanese Twitter community, focusing on the engagement patterns and perspectives of early and late engagers. The research employed a dual methodology, incorporating both quantitative and qualitative analyses to understand the nature of the discussions.

The quantitative analysis revealed distinct conversational focuses between early and late engagers. Early engagers emphasized forward-looking and speculative topics, highlighting the technological advancements and potential transformative impact of ChatGPT. In contrast, late engagers engaged more with contemporary topics, focusing on optimizing existing AI capabilities and considering their limitations.

The qualitative analysis delved deeper into the discussions and measured the breadth of perspectives within topics between the two groups. A weak correlation was found between the volume of tweets and the range of discussed topics in both groups. We also found that early engagers tended to concentrate on a more limited range of perspectives, while late engagers exhibited a broader range of viewpoints, even reducing the bias of the tweet volume. This finding emphasized the importance of identifying semantic bias and understanding the content and nuances of discussions beyond the volume of tweets.

The dual quantitative and qualitative methodology employed in this study is versatile and applicable to studying engagement patterns within other user groups or beyond the context of ChatGPT. The insights gained from this research contribute to a better understanding of public discourse surrounding emerging technologies and the communication styles of different user groups.

Future studies can extend this research in several ways.
Firstly, by utilizing the proposed methodology, we will be able to extract topics where semantic polarization is occurring and  determine the existence of conflicting opinions within those topics. Secondly, exploring engagement patterns in other community clustering methods such as network-based clustering methods can shed light on the influence of community structures on discussion dynamics. Additionally, grouping tweets based on their popularity, distinguishing between well-retweeted and rarely retweeted tweets, may reveal how only some perspectives are seen by many people. Furthermore, tracking the evolution of these engagement patterns over time can offer valuable insights into the development and maturation of user interactions and public discourse surrounding emerging technologies. By addressing these areas in future research, we can achieve a more comprehensive understanding of the complex dynamics involved in discussions about emerging technologies and their societal impact.

\bibliographystyle{IEEEtran}
\bibliography{bibliography}

\begin{thebibliography}{10}
\providecommand{\url}[1]{#1}
\csname url@samestyle\endcsname
\providecommand{\newblock}{\relax}
\providecommand{\bibinfo}[2]{#2}
\providecommand{\BIBentrySTDinterwordspacing}{\spaceskip=0pt\relax}
\providecommand{\BIBentryALTinterwordstretchfactor}{4}
\providecommand{\BIBentryALTinterwordspacing}{\spaceskip=\fontdimen2\font plus
\BIBentryALTinterwordstretchfactor\fontdimen3\font minus
  \fontdimen4\font\relax}
\providecommand{\BIBforeignlanguage}[2]{{%
\expandafter\ifx\csname l@#1\endcsname\relax
\typeout{** WARNING: IEEEtran.bst: No hyphenation pattern has been}%
\typeout{** loaded for the language `#1'. Using the pattern for}%
\typeout{** the default language instead.}%
\else
\language=\csname l@#1\endcsname
\fi
#2}}
\providecommand{\BIBdecl}{\relax}
\BIBdecl

\bibitem{Leiter2023ChatGPTAM}
C.~Leiter, R.~Zhang, Y.~Chen, J.~Belouadi, D.~Larionov, V.~Fresen, and S.~Eger,
  ``Chatgpt: A meta-analysis after 2.5 months,'' \emph{ArXiv}, vol.
  abs/2302.13795, 2023.

\bibitem{UBHD2028615}
E.~M. Rogers, \emph{\BIBforeignlanguage{eng}{Diffusion of innovations}},
  5th~ed.\hskip 1em plus 0.5em minus 0.4em\relax New York, NY [u.a.]: Free
  Press, 08 2003.

\bibitem{Haque}
M.~U. Haque, I.~Dharmadasa, Z.~T. Sworna, R.~Rajapakse, and H.~Ahmad, ``"i
  think this is the most disruptive technology": Exploring sentiments of
  chatgpt early adopters using twitter data,'' 12 2022.

\bibitem{Taecharungroj2023WhatCC}
V.~Taecharungroj, ``"what can chatgpt do?" analyzing early reactions to the
  innovative ai chatbot on twitter,'' \emph{Big Data Cogn. Comput.}, vol.~7,
  p.~35, 2023.

\bibitem{leiter}
C.~Leiter, R.~Zhang, Y.~Chen, J.~Belouadi, D.~Larionov, V.~Fresen, and S.~Eger,
  ``Chatgpt: A meta-analysis after 2.5 months,'' 02 2023.

\bibitem{antypas-etal-2022-twitter}
\BIBentryALTinterwordspacing
D.~Antypas, A.~Ushio, J.~Camacho-Collados, V.~Silva, L.~Neves, and F.~Barbieri,
  ``{T}witter topic classification,'' in \emph{Proceedings of the 29th
  International Conference on Computational Linguistics}.\hskip 1em plus 0.5em
  minus 0.4em\relax Gyeongju, Republic of Korea: International Committee on
  Computational Linguistics, Oct. 2022, pp. 3386--3400. [Online]. Available:
  \url{https://aclanthology.org/2022.coling-1.299}
\BIBentrySTDinterwordspacing

\bibitem{10.5555/3157382.3157584}
T.~Bolukbasi, K.-W. Chang, J.~Zou, V.~Saligrama, and A.~Kalai, ``Man is to
  computer programmer as woman is to homemaker? debiasing word embeddings,'' in
  \emph{Proceedings of the 30th International Conference on Neural Information
  Processing Systems}, ser. NIPS'16.\hskip 1em plus 0.5em minus 0.4em\relax Red
  Hook, NY, USA: Curran Associates Inc., 2016, p. 4356–4364.

\bibitem{944937}
\BIBentryALTinterwordspacing
D.~M. Blei, A.~Y. Ng, and M.~I. Jordan, ``Latent dirichlet allocation,''
  \emph{J. Mach. Learn. Res.}, vol.~3, pp. 993--1022, 2003. [Online].
  Available: \url{http://portal.acm.org/citation.cfm?id=944937}
\BIBentrySTDinterwordspacing

\bibitem{grootendorst2022bertopic}
M.~Grootendorst, ``Bertopic: Neural topic modeling with a class-based tf-idf
  procedure,'' \emph{arXiv preprint arXiv:2203.05794}, 2022.

\bibitem{zhang-etal-2022-neural}
\BIBentryALTinterwordspacing
Z.~Zhang, M.~Fang, L.~Chen, and M.~R. Namazi~Rad, ``Is neural topic modelling
  better than clustering? an empirical study on clustering with contextual
  embeddings for topics,'' in \emph{Proceedings of the 2022 Conference of the
  North American Chapter of the Association for Computational Linguistics:
  Human Language Technologies}.\hskip 1em plus 0.5em minus 0.4em\relax Seattle,
  United States: Association for Computational Linguistics, Jul. 2022, pp.
  3886--3893. [Online]. Available:
  \url{https://aclanthology.org/2022.naacl-main.285}
\BIBentrySTDinterwordspacing

\bibitem{Aggarwal01onthe}
C.~C. Aggarwal, A.~Hinneburg, and D.~A. Keim, ``On the surprising behavior of
  distance metrics in high dimensional space,'' in \emph{Lecture Notes in
  Computer Science}.\hskip 1em plus 0.5em minus 0.4em\relax Springer, 2001, pp.
  420--434.

\bibitem{2018arXivUMAP}
L.~{McInnes}, J.~{Healy}, and J.~{Melville}, ``{UMAP: Uniform Manifold
  Approximation and Projection for Dimension Reduction},'' \emph{ArXiv
  e-prints}, Feb. 2018.

\bibitem{MoulaviJCZS14}
\BIBentryALTinterwordspacing
D.~Moulavi, P.~A. Jaskowiak, R.~J. G.~B. Campello, A.~Zimek, and J.~Sander,
  ``Density-based clustering validation.'' in \emph{SDM}, M.~J. Zaki,
  Z.~Obradovic, P.-N. Tan, A.~Banerjee, C.~Kamath, and S.~Parthasarathy,
  Eds.\hskip 1em plus 0.5em minus 0.4em\relax SIAM, 2014, pp. 839--847.
  [Online]. Available: \url{http://dblp.uni-trier.de/db/conf/sdm/sdm2014.html}
\BIBentrySTDinterwordspacing

\bibitem{10.1145/3209581}
\BIBentryALTinterwordspacing
R.~Baeza-Yates, ``Bias on the web,'' \emph{Commun. ACM}, vol.~61, no.~6, p.
  54–61, may 2018. [Online]. Available: \url{https://doi.org/10.1145/3209581}
\BIBentrySTDinterwordspacing

\bibitem{FugledeTopsoe}
B.~Fuglede and F.~Topsoe, ``{Jensen-Shannon divergence and Hilbert space
  embedding},'' in \emph{IEEE International Symposium on Information Theory},
  2004, pp. 31--31.

\bibitem{10.1006/jmva.1997.1659}
\BIBentryALTinterwordspacing
C.~O. Wu, ``A cross-validation bandwidth choice for kernel density estimates
  with selection biased data,'' \emph{J. Multivar. Anal.}, vol.~61, no.~1, p.
  38–60, apr 1997. [Online]. Available:
  \url{https://doi.org/10.1006/jmva.1997.1659}
\BIBentrySTDinterwordspacing

\end{thebibliography}

\end{document}